\documentclass[sigconf]{acmart}

\AtBeginDocument{%
  \providecommand\BibTeX{{%
    \normalfont B\kern-0.5em{\scshape i\kern-0.25em b}\kern-0.8em\TeX}}}

\acmConference[ARXIV]{}{Jan 25, 2024}

\usepackage{listings}
\usepackage{xcolor}
\usepackage{makecell}

\begin{document}

\title{Designing Silicon Brains using LLM: Leveraging ChatGPT for Automated Description of a Spiking Neuron Array}

\author{Mike Tomlinson}
\email{mtomlin5@jh.edu}
\affiliation{%
  \institution{Johns Hopkins University}
  \streetaddress{3400 N Charles St, Baltimore, MD 21218}
  \city{Baltimore}
  \state{MD}
  \country{USA}
  \postcode{21218}
}

\author{Joe Li}
\email{qli67@jh.edu}
\affiliation{%
  \institution{Johns Hopkins University}
  \streetaddress{3400 N Charles St, Baltimore, MD 21218}
  \city{Baltimore}
  \state{MD}
  \country{USA}
  \postcode{21218}
}

\author{Andreas Andreou}
\email{andreou@jhu.edu}
\affiliation{%
  \institution{Johns Hopkins University}
  \streetaddress{3400 N Charles St, Baltimore, MD 21218}
  \city{Baltimore}
  \state{MD}
  \country{USA}
  \postcode{21218}
}

\renewcommand{\shortauthors}{Tomlinson and Li, et. al}

\begin{abstract}
Large language models (LLMs) have made headlines for synthesizing correct-sounding responses to a variety of prompts, including code generation. In this paper, we present the prompts used to guide ChatGPT4 to produce a synthesizable and functional verilog description for the entirety of a programmable Spiking Neuron Array ASIC. This design flow showcases the current state of using ChatGPT4 for natural language driven hardware design. The AI-generated design was verified in simulation using handcrafted testbenches and has been submitted for fabrication in Skywater 130nm through Tiny Tapeout 5 using an open-source EDA flow.

\end{abstract}


\maketitle

\section{Introduction}
Over the last three decades, advances in CMOS technology and CAD tools have led to advances in processor technology that in turn fed research into design and automation tools that enable the sophisticated System On Chip for general computing and AI. Verilog and VHDL, both released in the 1980s, have become standard synthesis tools in digital design. These tools allow the writer to describe behavioral functionality that can be directly mapped to digital standard cells and physical layout generation through place and route. Verilog and VHDL are the schematic entry point into modern CAD tools. Writing and maintaining code in Verilog and VHDL introduces significant overhead as these abstract design at at a rather low level. There are a number of projects trying to address this with a range of adoption and commercial support. These projects include efforts such as Chisel\cite{bachrach_chisel_2012} and High Level Synthesis (with specific tools from Cadence, Vivado, and Synopsys). Overall, the general trend in these methods is to move towards a higher level language that can then be used to generate VHDL or Verilog.

In November of 2023, OpenAI's ChatGPT, a LLM captured the attention of users and business alike because it offered a simple but powerful interface to LLMs for performing generative AI tasks. This interactive interface to LLMs is capable of executing a variety of tasks such as writing prose and generating code. This model has shown to be effective at generating python, albeit with problems of attention span and adaptability \cite{kashefi_chatgpt_2023}, \cite{tian_is_2023}. 

Recent works addressing LLM assisted hardware design include a LLM based optimization framework that integrates ChatGPT with existing EDA tools. In the work by \cite{chang_chipgpt_2023}, authors employ LLM tools to implement several simple modules. For each implementation, power, performance, and area are compared to modules implemented with ChatGPT alone, Xilinx HLS, and Chisel. Other research explores a different set of simple blocks, including a simple microprocessor with a ChatGPT-defined ISA \cite{blocklove_chip-chat_2023}. Similarly, Yang et al. investigate ChatGPT's effectiveness for systolic arrays and ML accelerators \cite{yang_new_2023}.

\subsection{Contributions}
In this paper we explore the use of generative AI and ChatGPT (version 4 is used in this work) \cite{OpenAI2023} to design a hardware system, namely a spiking neural network chip, a neuromorphic electronic system~\cite{Mead1990}, for hardware AI inference~\cite{Sanni2019}. Our effort differs from the current research in the use of LLMs for CAD by focusing on digital spiking neurons~\cite{Cassidy2013}, an unconventional computing architecture, and by emphasizing complete system design. We document the steps taken to go from a conversational design description to a functional and synthesizable Verilog description of a programmable array of spiking neurons. The final AI-generated HDL has a standardized interface, SPI, and multiple levels of hierarchy. This work represents one of the first ASICs synthesized entirely from natural conversational language. By documenting this process, we hope to showcase the current state of using LLMs as a higher level, conversational, alternative to handcrafted HDL.

\definecolor{codegreen}{rgb}{0,0.6,0}
\definecolor{codegray}{rgb}{0.5,0.5,0.5}
\definecolor{codepurple}{rgb}{0.58,0,0.82}
\definecolor{backcolour}{rgb}{0.95,0.95,0.92}

\lstdefinestyle{mystyle}{
    basicstyle=\scriptsize, 
    commentstyle=\color{codegreen},
    keywordstyle=\color{magenta},
    numberstyle=\tiny\color{codegray},
    stringstyle=\color{codepurple},
    basicstyle=\ttfamily\scriptsize,
    breakatwhitespace=false,         
    breaklines=true,                 
    captionpos=b,                    
    keepspaces=true,                 
    numbers=left,                    
    numbersep=5pt,                  
    showspaces=false,                
    showstringspaces=false,
    showtabs=false,                  
    tabsize=2
}

\lstset{style=mystyle}

\section{Natural Language Hardware Description} 

This work targets a neuromorphic design with a model trained on a large amount of public data. Only a small fraction of this data is likely to be related to neuromorphic engineering and an even smaller fraction to neuromorphic Verilog. Table \ref{tab:popularity} gives an idea of the volume of training code available for these topics. The table lists relevant keywords and the corresponding number of matching publicly available repositories on Github. There are hundreds of spiking neuron related repositories, but around 2 orders magnitude fewer dealing specifically with Verilog. 

\begin{table}
  \caption{Number of Search Results on Github}
  \label{tab:popularity}
  \begin{tabular}{lcc}
    \toprule
    Search Term & Language & Number of Results \\
    \midrule
    "integrate and fire" & Verilog & 3 \\
    "integrate and fire" & Any & 750 \\
    "snn" & Verilog & 26 \\
    "snn" & Any & 2.4k \\
    "spiking neuron" & Verilog & 6 \\
    "spiking neuron" & Any  & 478 \\
    "spiking" & Verilog  & 29 \\
    "spiking" & Any  & 8.7k \\
    "alu" & Verilog  & 1.1k \\
    "accelerator" & Verilog & 339 \\
    "cnn" & Verilog  & 198 \\
    "spi" & Verilog  & 530 \\
  \end{tabular}
\end{table}

We start by prompting ChatGPT to generate a leaky integrate and fire (LIF) neuron. This module is then instantiated in a network module, where 2 layers of 3 neurons are instantiated in a fully connected fashion. We then prompt for an SPI peripheral that will be used to program the network parameters. Finally, we ask ChatGPT to synthesize a top-level module instantiating both the SPI and network modules. All transcripts and source code can be found on Github : 
\url{https://github.com/AndreouLab/tinytapeout_05_chatgpt_snn}

\subsection{The Neuron}
The LIF neuron model is one of the more popular spiking neuron models and one of the simplest. The basic LIF neuron~\cite{Cassidy2013} has one state variable, the membrane potential, $V_m$, and one output, $S$, a boolean variable indicating if a spike is generated or not. In discrete time, these variables can be modeled by the following equations:

\[ V_{m}[t] = \begin{cases} 
      max(I_{in}[t] + V_{m}[t-1] - I_{leak}, 0), & S[t] = 0 \\
      0, & S[t] = 1 \\
   \end{cases}
\]

\[ S[t] = V_{m}[t] > V_{th}[t]\]

Additional functionality can be added by including a refractory filter. After a spike occurs, $S[t]$ transitions from 0 to 1; $V_{m}[t]$ will remain at 0 for a fixed period of time. This can be thought of as an additional state variable. 

This discrete time approximation of the LIF neuron model naturally lends itself to digital circuitry. The membrane potential can be implemented with an accumulator, the refractory filter, a counter, and the spiking output with a comparator. With this in mind, we prompted ChatGPT4 with a high level request for a Verilog module for a spiking LIF neuron: 

\hfill \break
Prompt: \textit{Can you write a verilog module for a spiking leaky integrate and fire neuron? Please include a multibit current input and a refractory period.} \\
\hfill \break
ChatGPT: \textit{Sure, I’ll provide a basic Verilog module for a spiking leaky integrate-and-fire neuron. Here’s a simple version: } \\(Shown in Table \ref{tab:neurons})

\begin{table*}
  \caption{Initial and Complete Generated LIF Neuron Verilog}
  \label{tab:neurons}
  \begin{tabular}{p{0.45\linewidth} p{0.025\linewidth} p{0.45\linewidth} }
      \toprule
      Initial Module & & Complete Module \\ 
      \midrule
\begin{lstlisting}[language=verilog]
module leaky_integrate_fire_neuron (
    input clk,               // clock input
    input reset,             // asynchronous reset
    input [7:0] current,     // 8-bit current input
    output spike             // output spike signal
);
// Parameters
parameter THRESHOLD = 8'd255; // spiking threshold
parameter LEAK_RATE = 8'd1;   // decrease by 1 per cycle
parameter REFRAC_PERIOD = 32; // refractory period in cycles

// Internal states
reg [7:0] membrane_potential; // current value of the membrane potential
reg [5:0] refrac_counter = 6'd0; // refractory period counter
reg in_refrac = 0; // refractory period flag

// On every clock cycle
always @(posedge clk or posedge reset) begin
    if (reset) begin
        membrane_potential <= 8'd0;
        refrac_counter <= 6'd0;
        in_refrac <= 0;
    end else begin
        if(in_refrac) begin
            refrac_counter <= refrac_counter - 1'b1;
            if(refrac_counter == 6'd0) in_refrac <= 0;
        end else begin
            // update membrane potential based on current input and leak rate
            if (membrane_potential + current > THRESHOLD) begin
                membrane_potential <= 8'd0;
                spike <= 1;
                in_refrac <= 1;
                refrac_counter <= REFRAC_PERIOD;
            end else if (membrane_potential + current - LEAK_RATE < 8'd0) begin
                membrane_potential <= 8'd0;
            end else begin
                membrane_potential <= membrane_potential + current - LEAK_RATE;
            end
        end
    end
end

// Reset spike signal after one cycle
always @(posedge clk or posedge reset) begin
    if (reset) spike <= 0;
    else if (spike) spike <= 0;
end
endmodule
\end{lstlisting} & & 
\begin{lstlisting}[language=verilog]
module leaky_integrate_fire_neuron (
    input clk,                  // clock input
    input reset,                // asynchronous reset
    input [7:0] current,        // 8-bit current input
    input [7:0] THRESHOLD,      // spiking threshold
    input [7:0] LEAK_RATE,      // decrease by 1 per cycle
    input [7:0] REFRAC_PERIOD,  // refractory period in cycles
    output reg spike            // output spike signal
);
// Internal states
reg [7:0] membrane_potential = 8'd0; 
reg [7:0] refrac_counter = 8'd0; 
reg in_refrac = 0; 

// On every clock cycle
always @(posedge clk or posedge reset) begin
    if (reset) begin
        membrane_potential <= 8'd0;
        refrac_counter <= 8'd0;
        in_refrac <= 0;
        spike <= 0;
    end else begin
        spike <= 0; // By default, reset the spike signal          
        if(in_refrac) begin
            refrac_counter <= refrac_counter - 1'b1;
            if(refrac_counter == 8'd0) in_refrac <= 0;
        end else begin
            // Check for potential underflow
            if (membrane_potential < LEAK_RATE) begin
                membrane_potential <= current; // Directly set to current
            end else if (membrane_potential + current < membrane_potential) begin
                // Potential overflow detected, clamp to THRESHOLD
                membrane_potential <= THRESHOLD;
            end else begin
                membrane_potential <= membrane_potential + current - LEAK_RATE;
            end
            
            // Check for spiking
            if (membrane_potential >= THRESHOLD) begin
                membrane_potential <= 8'd0;
                spike <= 1;
                in_refrac <= 1;
                refrac_counter <= REFRAC_PERIOD;
            end 
        end
    end
end
endmodule
\end{lstlisting}  \\
      \bottomrule
  \end{tabular}
\end{table*}
\hfill \break
After a cursory look at the code, it may seem very impressive! But, after careful consideration and simulation, a number of problems can be found. 

\begin{enumerate}
  \item \textbf{Line 5}: syntax error, spike is declared as a wire (by default), it needs to explicitly declared as a register
  \item \textbf{Lines 27-38}: logical error, update logic does not account for overflow
  \item \textbf{Lines 34}: logical error, faulty logic for checking for an underflow (line 34)
  \item \textbf{Lines 18, 45}: structural error, spike is multiply driven (used in two processes)
\end{enumerate}

These problems were all fixable through further prompting. Issues 1 and 4 each required only one additional prompt to correct. The logical errors, 2 and 3, were considerably more difficult, requiring several iterations to correctly handle overflow, underflow, and working with unsigned numbers. 

Once the code was functional and synthesizable, we prompted ChatGPT to make the neuron parameters (threshold, leak rate, and refractory period) programmable. ChatGPT was able to do this without error, deleting the parameter declarations and adding input ports to drive the parameter values. We also increased the bit width of the refractory period from 6 to 8 bits through prompting. The complete Verilog for this module is shown in Table \ref{tab:neurons}. 

ChatGPT had little trouble understanding the basic concept of the LIF neuron and making specific modifications when given a detailed prompt, but struggled to provide complete modules without syntax errors and showed conceptual misunderstanding when working with limited precision unsigned numbers. For an experienced digital designer these problems are relatively easy to fix, but for a user without the same background, these problems may pose considerable difficulty. Requiring extensive code modifications heavily limits the ability for ChatGPT to act as a full natural language abstraction.

\subsection{Network}
\begin{figure}[h]
  \centering
  \includegraphics[width=0.9\linewidth]{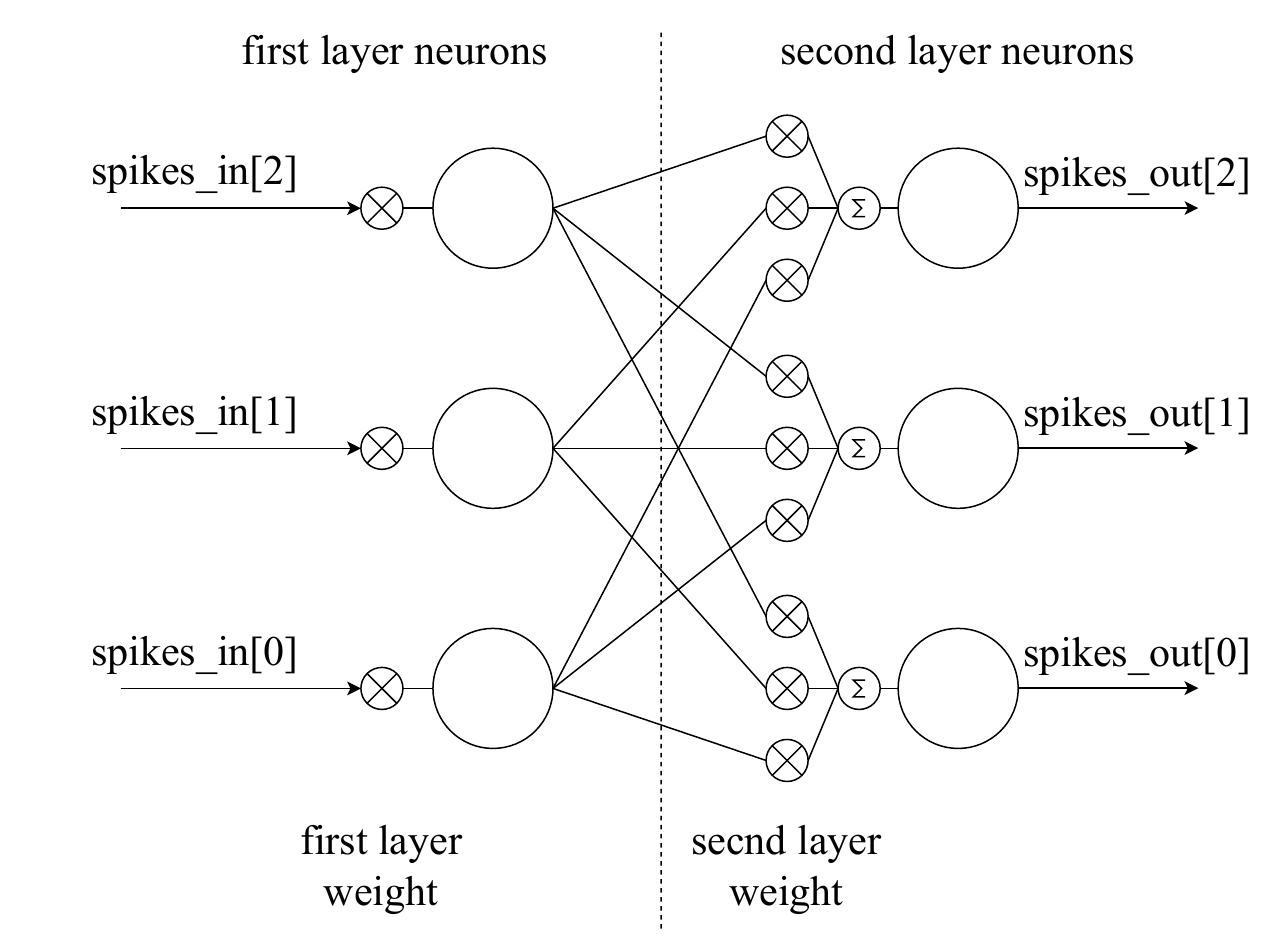}
  \fontsize{12}{10}\selectfont 
  \caption{Two layer fully connected neuron network.}
  \label{fig:network_diagram}
\end{figure}

For the network, we wanted to implement something relatively simple. We decided on a 2 layer network with 3 neurons per layer and fully programmable weights. A diagram of the envisioned network is shown in Fig. \ref{fig:network_diagram}.

Each circle represents one LIF neuron. For a given neuron, the conversion from spikes to current can be defined as the sum of incoming spikes weighted by the strength of the synaptic connection. Mathematically, the input current for each neuron is given by:
\[ I_{in}[t] = \sum_{i=1}^{N} w_ix_i[t]\]

Where $I_{in}$ is the neuron input current, $x_i$ is the incoming spike from neuron i in the previous layer, $w_i$ is the programmed weight for the ith input to this neuron, and N is the total number of synaptic connections to the previous layer. To describe this module in Verilog, an experienced designer might consider using two generate statements, one per layer, and some combinational logic to calculate the input currents. We opened a new chat, provided a new copy of the neuron module and then gave the following prompt:

\hfill \break
Prompt: \textit{Now that we have a working neuron, can you create a new module that instantiates a network of neurons with 2 layers in a fully connected fashion? Please connect them with programmable synapses. }
\hfill \break

The generated code is included in Table \ref{tab:network}. For this module, there are a number of apparent issues. 

\begin{enumerate}
  \item \textbf{Line 7}: syntax error, parameter N is used but never declared
  \item \textbf{Lines 7, 9}: syntax error, ports, input\_currents and weights, are declared as 2d arrays, which is invalid in Verilog (although valid in systemVerilog)
  \item \textbf{Line 8}: syntax error, similar to the previous issue, spikes is declared as an unpacked array, which is not allowed for ports in verilog (although valid in systemVerilog)
  \item \textbf{Lines 13-21}: syntax error, the instantiation of the neuron is not valid syntax, no generate variable is used, nor are variables used to index into the current and spike connections
  \item \textbf{Lines 40-48}: syntax error, issue 4 also applies to the second set of instantiations
  \item \textbf{Lines 25-37}: syntax error, spike variable access is syntactically incorrect.
\end{enumerate}

Issues 1 and 3 were relatively easy to fix, each only requiring one additional prompt. Issue 2 was more difficult. Eventually, the issue was resolved by suggesting that ChatGPT flatten the 2D ports into 1D arrays. This required the prompter to be familiar with Verilog in order to provide the solution, limiting the use of this process to already experienced engineers. 

After flattening the array, ChatGPT created the following code to reassign the flattened array to an internal 2D variable. This code has a fundamental problem. An initial block is used when continuous assignment is desired. This error causes weights to be initialized, but never updated after initialization, resulting in a module that cannot be programmed. Prompting ChatGPT to fix this issue caused it to then declare spikes (Line 8) as type reg, introducing another syntax error (spikes is driven through a port connection). In fixing one issue, two others were introduced. 

\begin{lstlisting}[language=Verilog] 
reg [7:0] weights[2:0][2:0];
integer i, j;
initial begin
    for (i = 0; i < 3; i = i + 1) begin
        for (j = 0; j < 3; j = j + 1) begin
            weights[i][j] = weights_flat[(i*3 + j)*8 +: 8];
        end
    end
end     
); \end{lstlisting} 

Issue 4 was also difficult to resolve, requiring 7 additional prompts. We requested ChatGPT explicitly declare each neuron separately and then concatenate those instantiations back together using a generate statement. While working on this part of the code, ChatGPT again introduced syntax errors in other parts of the code. At one point, replacing \textit{begin} in the code snippet above with an open bracket while keeping the corresponding end statement, mixing C style coding and Verilog. Similar to the neuron module, these issues required knowledge of Verilog to identify, debug, and resolve through prompting, reducing the level of abstraction.

Once the errors were corrected, we used ChatGPT to make design revisions to the generated code. For this module we added a register file to store the neuron parameters (shared for all neurons), and the network weights. We first gave the following prompt:

\hfill \break
Prompt: \textit{Ok, can you please provide a programmable register file that stores all the parameters to our network}
\hfill \break

This prompt produced a reasonable looking register file, that stored the neuron parameters for each neuron (threshold, leak rate, and refractory period). The produced code was syntactically correct, but not exactly what we were asking for. This highlights both an impressive and problematic aspect of this technology. The previous prompt was ambiguous. It is not clear what parameters we wanted stored; and rather than asking for clarification, ChatGPT assumes we want to store independent parameters for each neuron, instead of storing the weights and a single set of neuron parameters to be shared. This was an easy clarification to make. The next prompt resulted in an almost correct module with our desired functionality and only a minor syntax error, confusing systemVerilog and Verilog coding styles.  

\hfill \break
Prompt: \textit{Can you include weights as a parameter and let's say both layers use the same parameters }
\hfill \break

Again, this prompt was somewhat vague. From the structure of the sentence, we can infer that the intent is to store independent values for each weight and a single common copy of the neuron parameters. ChatGPT was able to make the same inference. Once this module was
functionally correct, we prompted ChatGPT to merge this functionality into the network module. This was done without error. 

\begin{table*}
  \caption{Initial Generated Network and SPI modules}
  \label{tab:network}
  \begin{tabular}{p{0.45\linewidth} p{0.025\linewidth} p{0.45\linewidth} }
      \toprule
      Network Module & & SPI Module \\ 
      \midrule
\begin{lstlisting}[language=verilog]
module two_layer_neural_network (
    input clk,                      // clock input
    input reset,                    // asynchronous reset
    input [7:0] THRESHOLD,          // spiking threshold
    input [7:0] LEAK_RATE,          // decrease by 1 per cycle
    input [5:0] REFRAC_PERIOD,      // refractory period in cycles
    input [7:0] input_currents[N],  // Input currents for N neurons in the first layer
    output spikes[M],               // Spike outputs for M neurons in the second layer
    input [7:0] weights[N][M]       // Programmable synapse weights
);

    // First layer of neurons
    leaky_integrate_fire_neuron first_layer[N](
        .clk(clk),
        .reset(reset),
        .THRESHOLD(THRESHOLD),
        .LEAK_RATE(LEAK_RATE),
        .REFRAC_PERIOD(REFRAC_PERIOD),
        .current(input_currents),
        .spike() // We'll compute this shortly
    );

    // Logic to compute effective current for second layer neurons based on spikes and weights
    reg [7:0] second_layer_currents[M];
    genvar i, j;
    generate
        for (i = 0; i < M; i = i + 1) begin : compute_second_layer_current
            always @(posedge clk) begin
                second_layer_currents[i] = 0;
                for (j = 0; j < N; j = j + 1) begin
                    if (first_layer[j].spike) begin
                        second_layer_currents[i] = second_layer_currents[i] + weights[j][i];
                    end
                end
            end
        end
    endgenerate

    // Second layer of neurons
    leaky_integrate_fire_neuron second_layer[M](
        .clk(clk),
        .reset(reset),
        .THRESHOLD(THRESHOLD),
        .LEAK_RATE(LEAK_RATE),
        .REFRAC_PERIOD(REFRAC_PERIOD),
        .current(second_layer_currents),
        .spike(spikes)
    );

endmodule
\end{lstlisting} & & 
\begin{lstlisting}[language=verilog]
module spi_slave (
    input clk,         // System Clock
    input reset,       // System Reset
    input sclk,        // SPI Clock
    input cs_n,        // SPI Chip Select (Active Low)
    input mosi,        // SPI Master Out Slave In
    output reg miso,   // SPI Master In Slave Out
    output reg [3:0] addr_out,      
    output reg [7:0] data_out,      
    output reg write_enable         
);
    // SPI state definitions
    typedef enum {IDLE, TRANSFER, COMPLETE} spi_state;
    reg [1:0] state, next_state;
    reg [15:0] shift_reg; 
    reg [7:0] addr_reg;  
    integer bit_count;
    // SPI state machine
    always @(posedge clk or posedge reset) begin
        if (reset) begin
            state <= IDLE;
        end else begin
            state <= next_state;
        end
    end
    // SPI logic
    always @(posedge clk) begin
        case (state)
            IDLE: begin
                if (!cs_n) begin
                    bit_count <= 15;  // Reset bit count
                    next_state = TRANSFER;
                end
            end
            TRANSFER: begin
                shift_reg[bit_count] <= mosi;
                if (bit_count == 0) begin
                    next_state = COMPLETE;
                end else begin
                    bit_count <= bit_count - 1;
                end
            end
            COMPLETE: begin
                if (cs_n) begin
                    next_state = IDLE;
                    addr_reg <= shift_reg[15:8]; 
                    data_out <= shift_reg[7:0];  
                    write_enable <= 1;
                end
            end
        endcase
    end
    // MISO signal assignment
    always @(posedge clk) begin
        if (state == TRANSFER) begin
            miso <= shift_reg[15];
        end
    end
    always @(*) begin
        addr_out = addr_reg;
        if (state != COMPLETE) begin
            write_enable = 0;
        end
    end
endmodule
\end{lstlisting}  \\
      \bottomrule
  \end{tabular}
\end{table*}

\subsection{Serial Peripheral Interface (SPI)} 
We then prompted ChatGPT to create an SPI peripheral module, allowing us to program the network parameters and weights over a standard interface.

\hfill \break
\noindent Prompt: \textit{OK, I want you to now create a SPI interface to communicate with the network module above.}
\newline

The generated code is presented in Table \ref{tab:network}, some comments were removed to reduce the size of the code. This module also has a number of subtle errors. 

\begin{enumerate}
    \item \textbf{Line 13}: syntax error, use of the SystemVerilog enum feature which is not supported in Verilog
    \item \textbf{Lines 27-52}: logical errors in SPI finite state machine
    \item \textbf{Line 48, 62}: syntax error, write\_enable is multiply driven
\end{enumerate}

Errors 1 and 3 are similar to errors seen previously. The first is another mixup with SystemVerilog syntax and the third is a multiply driven net. Both of these issues were resolved with a single follow-up prompt addressing each. These consistent mistakes highlight ChatGPT's lack of familiarity with Verilog.

Error 2 also showcases a lack of experience with Verilog and with SPI. This code section confuses one segment and two segment coding styles. Lines 19-25 are typical for a two segment style, but inconsistent with lines 27-52 where \textit{next\_state} is assigned sequentially, as if \textit{next\_state} is the state variable in a one segment process. The state logic also has no dependence on \textit{sclk}, even though it created a port for this signal. \textit{sclk} is intended to be the SPI clock, which synchronizes the controller and peripheral. This signal is crucial to writing a correct SPI state machine. 

Additionally, the state machine doesn't come out of reset. When reset is applied, state is driven to IDLE and next state is undriven. When reset is deasserted, the current state will go to whatever \textit{next\_state} was at the time of applying reset. These issues were all eventually resolved through subsequent prompts.

\subsection{Top Module}
The top module needs to instantiate both the SPI module and the network module, appropriately connecting internal signals and declaring input and output ports. In the same chat, the following prompt was used:

\hfill \break
\noindent Prompt: \textit{Can you create a top file to connect this spi module with the network module?}
\newline

The generated code had only one error. The intermediate signals between the modules were declared as reg instead of wire. This was an easy fix. After one more prompt, the top module was complete. 

\section{Tiny Tapeout: Implementation}
This design was submitted to TinyTapeout 5, a multi-project die effort for Skywater 130nm through efabless. The flow is intended to handle most of the intricacies of digital implementation, exposing only a minimal set of configuration options in a user level script. This flow uses yosys \cite{wolf_yosys_nodate} for synthesis, which introduced an interesting complication. Through an iterative debugging process we determined the following code produced a multiply driven net error. 

\begin{lstlisting}[language=verilog]
always @(posedge clk or posedge reset) begin
    if (reset) begin
        ...
        for (idx1 = 0; idx1 < 3; idx1 = idx1 + 1) begin
           FIRST_LAYER_WEIGHTS[idx1] <= 8'd0;
           ...
        end
    end else if (write_enable) begin
        ...
    end
end

always @(*) begin
    for (idx1 = 0; idx1 < 3; idx1 = idx1 + 1) begin
        input_currents[idx1] = spikes_in[idx1] ? FIRST_LAYER_WEIGHTS[idx1] : 8'd0;
    end
end

Error: : Yosys checks have failed: Encountered check error:
Warning: Drivers conflicting with a constant 1'0 driver:
    port Q[2] of cell $procdff$614 ($dff)
\end{lstlisting}

After extensive debugging, we found that this error could be resolved by removing the duplicate loop variables in the code above. Once identified, simply asking ChatGPT to create new loop variables resolved the issue. Final simulations using the gate level netlist were conducted through the TinyTapeout flow. The design occupies 33\% of 320um x 200um and is expected to be fabricated by Summer 2024. 


\section{Conclusion}

This paper explores the use of ChatGPT to convert from natural language to functionally correct and synthesizable Verilog. We successfully use natural language entry to generate a complete HDL description of a programmable spiking neuron array, ready for implementation. It is clear technologies like ChatGPT have the potential to increase design efficiency, correctly producing simple modules, quickly generating foundational code from scratch and offering near instantaneous, accurate modifications of existing code when prompted detailed instructions. However, the current quality of ChatGPT's output often falls short. ChatGPT's responses tend to frequently include some form of error, either syntactically or logically. ChatGPT also confidently demonstrates ignorance of more advanced concepts, leading to potentially obfuscated bugs, increasing the difficulty of verification. These problems compound, placing a significant burden on the prompter. If the prompter knows the solution, they can guide ChatGPT to the answer; but without that knowledge, it can be difficult to use this technology as a tool for abstracting Verilog description. Overall, our findings suggest that natural language to Verilog synthesis has potential; but in its current form, it leaves much to be desired. 

\begin{acks}
This work was supported by NSF Grant 2020624 AccelNet:Accelerating Research on Neuromorphic Perception, Action, and Cognition through the Telluride Workshop on Neuromorphic Cognition Engineering, NSF Grant 2332166 RCN-SC: Research Coordination Network for Neuromorphic Integrated Circuits and NSF Grant 2223725 EFRI BRAID: Using Proto-Object Based Saliency Inspired By Cortical
Local Circuits to Limit the Hypothesis Space for Deep Learning Models.

\end{acks}

\bibliographystyle{ACM-Reference-Format}
\bibliography{chatgpt,aga}










\end{document}